\documentstyle[11pt,A4]{article}

\begin{document}

\title{{\bf  Random sequential adsorption on a dashed line}}
 \author{{\sc B. Bonnier, Y. Leroyer and E. Pommiers}\\[1cm]
{\em Laboratoire de Physique
Th\'eorique} \\ {\em CNRS, Unit\'e Associ\'ee 764}\\
{\em Universit\'e Bordeaux I}\\
{\em 19 rue du Solarium, 33175 Gradignan Cedex, France}}
\date{}
\begin{titlepage}

\maketitle
\thispagestyle{empty}
\begin{abstract}
We study  analytically and numerically  a
model of  random sequential adsorption (RSA) of segments on a line, subject
to some constraints suggested by two kinds of physical situations:
{\em (i)} deposition of dimers on a lattice where the sites have
a spatial extension; {\em (ii)} deposition of extended particles which must
overlap one (or several) adsorbing sites on the substrate.
Both systems involve discrete and continuous degrees
of freedom, and, in one dimension, are equivalent to our model, which
depends on one length parameter.
When this parameter is varied, the model  interpolates
between a variety of known situations~: monomers on a lattice, ``car-parking"
problem, dimers on a lattice. An  analysis of the long-time
behaviour of the coverage as a function of the parameter
exhibits an anomalous $1/t^2$ approach  to the jamming limit at the
transition point between
the fast exponential kinetics, characteristic of the lattice model, and
the $1/t$ law of the continuous one.
\end{abstract}
\vfill
LPTB 94-12\\
October 1994
e-mail : {\bf Leroyer@frcpn11.in2p3.fr}
\end{titlepage}

\section{Introduction}
The model of random sequential adsorption describes deposition processes
in which desorption is negligible and surface diffusion very slow on the
experimental time scale. Particles land successively and randomly
on the surface; due to the geometrical exclusion effect, if an incoming
particle
overlaps a previously deposited one, it is rejected. This model applies
to many physical situations such as adsorption of latex balls, proteins or
chemisorption at low temperature \cite{Evans}.
The substrate is either a lattice
or a continuous surface, depending on the size of the particles relative
to the microscopic scale. Many versions of the model have been studied
to adapt it to various physical situations, but all share
a common universal behaviour for the long time approach to the asymptotic
coverage, depending on the discrete or continuous nature of the substrate.
As long as the minimum interval between two neighbouring particles on
the substrate remains non-zero, which is the case on a lattice, one can show
that the jamming limit is approached exponentially\cite{Evans,EvansNord}.
Conversely, on a continuum substrate, the kinetics follows a power-law decay
whose exponent depends on the number of degrees of
freedom per particle[3-11].
For instance, in the one dimensional case, where the model is exactly
soluble, we have
$$\theta_k(t)=\theta_k(\infty )-A(k)e^{-\phi\; t}$$
for the deposition of $k$-mers on a lattice with flux $\phi$, whereas
$$\theta (t)=\theta (\infty )-A/t$$
for the continuous ``car-parking'' problem\cite{Renyi}.\\

In fact, at the mesoscopic scale, many physical situations involve
both continuous and discrete degrees of freedom. For example, with the
recent advances in nano-technologies, it is conceivable to realize a
lattice of small gold droplets deposited on a silicon surface
\cite{japs}.
The adsorption on this substrate of elongated particles (polymers)
whose length is of the order of the distance between two
gold dots can be
modelized by the adsorption of dimers \cite{Aime}. However, since the
size of the gold dots (typically some nanometers of radius) is not
negligible with respect to the length of the  particles, there will
be a continuous range of positions for the dimer to be fixed on two
neighbouring dots.

Alternatively, large particles like proteins or enzymes can be absorbed
on a latticized substrate.
Such a physical situation is described in ref.\cite{Tarjus2} and modelized
in the following way~:
adsorbing sites are randomly or regularly
disposed on a continuum substrate. Extended particles, represented by disks,
land on the surface but remain stuck only if they overlap one (or several)
adsorbing site(s). Here again, the deposition process is driven by the
discrete degree of freedom imposed by the location of the sites, and by
a continuous one associated with the position of the adsorbed particle
with respect to the site.\\

In this paper, we investigate a one-dimensional RSA model which involves both
discrete and continuous degrees of freedom. We show that
the two above-mentioned physical situations, when reduced
to a one-dimensional substrate, are equivalent to our model. In section 2,
we present the model and analyse its jamming limit using the result of
a numerical simulation. Section 3 is devoted to the analysis of
the long time behaviour of the coverage as derived from the master equations
of the model. Our results and the conclusions
are summarized in the section 4. The Appendices  contain the technical details
of the derivation of the master equations (Appendix A) and of their
solution (Appendix B).

\section{The model}
\subsection{Two equivalent one-dimensional models}

Consider the first example depicted in the introduction. It can be considered
as a generalisation of the lattice dimer model,
in which the lattice sites have a non zero extension which we set to unity.
In the deposition process, the dimer ends must
stick on this {\em discrete} set of {\em continuous} intervals, the situation
where a given interval contains both the end of one dimer and the origin
of another being now allowed (see figure 1(a)). Moreover,
the dimer length is no longer constrained to be equal to the lattice
spacing, but is
arbitrary within bounds compatible with these adsorbing rule, thereby
introducing a new length scale. We denote by $a$
the edge-to-edge distance between two neighbouring sites. The dependence
of the model on this parameter occurs
only in a trivial redefinition of the effective
flux of incoming particles. Therefore, the model is independent of
$a$, provided we impose the condition that
a deposited particle overlaps an inter-site interval.
The  length $\ell$ of the dimer lies in
the interval $[a,a+2]$. We set $\ell=a+r$, where $r\in [0,2]$ will
be the unique parameter of the model. In the following, we will refer
to this model as model (I).

Consider now the second example of the introduction where the adsorbing
sites are regularly disposed and the radius of the particles is
smaller than the inter-site distance. In one dimension, the substrate is
a regular lattice, the particle a segment of length $r<2$ lattice units,
and the adsorbing
rule, aside from the RSA ones, is that the segment must cover a site
(see figure 1(b)). We call this model, model (II).

If we constrain the deposited particle to overlap {\em only one}
adsorbing site,
it is obvious that model (II) is equivalent to the previous one taken in the
limit $a=0$, where the inter-site interval, reduced to a point, is
identified with the adsorbing site of model (II).
In the following, we will consider  the one dimensional model
from this point of view and from now on everything will refer to model (II).

\subsection{The jamming limit}
The constraint that a deposited particle must overlap one and only one
adsorbing site
just modifies the incoming flux $\phi$  by a multiplicative factor of
$r$ for $r<1$ and $2-r$ for $r>1$.
We define the occupancy rate $\theta$ as the fraction of
occupied sites. The covering of the whole substrate is $r\theta$.\\

 The jamming limit for this quantity, obtained from a numerical simulation,
is presented in figure 2, as a function of $r$.
Let us first analyse this curve qualitatively.  Clearly, for $r\leq 1/2$,
the adsorption of a particle on the
site $i$, regardless of its position on the site, cannot prevent the
deposition
on neighbouring sites. Therefore, each site will be occupied independently
of its neighbours, and
the model is completely equivalent to the deposition of monomers
on a lattice; we expect the asymptotic limit $\theta=1$ to be reached
exponentially fast.

Consider now for the site $i$ the most defavourable situation
in the case $r<1$, (fig. 3(a)),
where the site $i-1$ is occupied by the extreme left edge of a particle
 and the site $i+1$ by the
extreme right edge of another one. In this situation
the interval for a deposition at site $i$ is minimum and has extension
$2(1-r)$. The deposition is allowed if $2(1-r)>r \Longleftrightarrow r< 2/3$.
Therefore, for $r<2/3$,
we still expect a coverage of {\em all} the sites, which is
effectively observed in figure 2 where $\theta (t=\infty )=1$ up to $r=2/3$.
Furthermore, since in this $r$-interval the smallest target is of
non-zero extension, the kinetics must remain lattice-like and
the jamming limit is again approached  exponentially fast.

Conversely consider now the most favourable situation
in the case $r>1$ (fig. 3(b)), where the particle deposited on
the site $i-1$ has its extreme left edge close to site $i-2$.
The remaining  space to adsorb a particle on site $i$, has extension $3-r$.
If this interval is less than $r$ ($\Rightarrow r>3/2$), deposition
is impossible,
and the particle at site $i-1$ {\em effectively occupies two sites}
\footnote{This result is more clearly seen in the framework of model (I)}.
In this situation, for $r>3/2$, the model is
completely equivalent to the lattice dimer model \cite{Flory} and
one expects $\displaystyle{\theta (t=\infty )=\frac{1-e^{-2}}{2}}=0.43233...$
(see figure 2).

Between these two extreme cases, for $2/3\leq r\leq 3/2$, the asymptotic
occupancy rate $\theta (t=\infty )$ decreases continuously.
Furthermore, since the target intervals
for a particle deposition can be arbitrarily small, one expects a power law
dynamics, characteristic of a {\em continuous} model. The case $r=1$ is
special: the particle length matches exactly the inter-site distance and
the constraint for a landing particle to overlap one and only one
adsorbing site is automatically satisfied;
the discrete nature of the substrate no longer plays any role
 and we recover the ``car-parking" problem\cite{Renyi} with jamming limit
$\theta=0.747597...$.\\

\section{The long time behaviour of the coverage}
\subsection{The master equations}
We have derived the master equations for the time evolution of
the probability of finding at time $t$ a gap of given length. All the
technical details are given in appendix A. The result is that the
coverage $\theta (t)$ can be expressed in terms of a reduced probability
$P(x,t)$~:
\begin{eqnarray}
\theta (t) & = & \int_0^t dt'\; e^{rt'}\int_0^r dx\; P(x,t')\;P(r-x,t')
\qquad {\rm for}\; r\in ]0,1]
\label{thet1}\\
      & = & \int_0^t dt'\; e^{(2-r)t'}\int_{r-1}^1 dx\; P(x,t')\;P(r-x,t')
\qquad {\rm for}\; r\in [1,2[
\label{thet2}
\end{eqnarray}
where $P(x,t)$ is solution of the integral equation
\begin{eqnarray}
{\rm case}\; r\leq 1&&x\in[0,r]\nonumber\\
& & \nonumber\\
-\frac{\partial P(x,t)}{\partial t} & = &
xP(x,t)+\int_x^r P(x',t)dx'+
u(x+r-1)\; e^{-rt}\int_0^{x+r-1}P(x',t)dx' \label{ME12}\\
 & & \nonumber\\
{\rm case}\; r\geq 1&&x\in[r-1,1]\nonumber\\
& & \nonumber\\
-\frac{\partial P(x,t)}{\partial t} & = &
(x-r+1)P(x,t)+\int_x^1 P(x',t)dx'+ e^{-(2-r)t}\;
\int_{r-1}^{b(x)}P(x',t)dx' \label{ME22}
\end{eqnarray}
subject to the initial condition $P(x,0)=1$. In eq.(\ref{ME12}),
 $u(t)$ is the Heaviside step function and in the last integral of
eq.(\ref{ME22}), $b(x)={\rm Inf}\; \{x+r-1,1\}$. \\

These equations can be exactly solved in a few special cases,
corresponding to $r\leq 1/2$, $r=1$ and $r\geq 3/2$, which are
presented in the next section. For the generic case $1/2<r<3/2$ ($r\neq 1$),
we have devised an iterative construction of the general solution,
which is developed in appendix B and
which leads to the following result~:

Due to the last term in eqs(\ref{ME12},\ref{ME22}), $P(x,t)$ appears to be
piecewise
defined with respect to $x$, in successive intervals of length $(1-r)$
within the range $(0,r)$ for $r\leq 1$, or of length $(r-1)$
within the range $(r-1,2-r)$ for $r\geq 1$. The number of intervals thus
depends on the value of $r$ and is given by (see appendix A)~:
\begin{equation}
k = \left[\frac{r}{1-r}\right]+1\quad  {\rm for}\; 1/2< r< 1
\qquad
k = \left[\frac{2-r}{r-1}\right]+1\quad  {\rm for}\; 1< r<3/2
\label{kint1}
\end{equation}
where $[X]$ denotes the integer part of $X$.

The function $P_\ell (x,t)$
which coincides with $P(x,t)$ in the $\ell^{\rm th}$ interval is expressed
in terms of a unique $x$-independent function $q(t)$
\begin{eqnarray}
P_{\ell}(x,t) & = & q(t)+\int_0^t K_{\ell}(x,t|t')\; q(t')\; dt \qquad
2\leq \ell \leq k   \label{pl}\\
P_1(x,t) & \equiv & q(t)\nonumber
\end{eqnarray}
where the kernels $K_{\ell}(x,t|t')$ are constructed recursively.
The function $q(t)$ is itself solution of a linear integral equation
\begin{equation}
q(t) = u_0(t)\left\{ 1+\int_0^t \rho_k (t,t')\; q(t')\; dt'\right\} \label{qt}
\end{equation}
where  $u_0(t)$ is defined by
\begin{eqnarray*}
1/2<r<1 && u_0(t)=e^{-rt} \\
1<r<3/2 && u_0(t)= \exp\left[e^{-(2-r)t}-(2-r)t-1\right]
\end{eqnarray*}
and the kernels  $\rho_k$ are expressed in terms of $K_\ell$.\\

Even in the simplest case, $k=2$, we are unable to find  analytic
solutions of equation~(\ref{qt}).
However, from these expressions we can extract both
approximate solutions and the exact asymptotic behaviour for the dynamics
of the model, which allow us to understand the various regimes which
interpolate between the special cases depicted in section 3.2. \\
This is the object of the following sections.

\subsection{Special cases}
There are 3 special cases where the rate equations (\ref{ME12},\ref{ME22})
reduce to a simple form which is exactly solvable.
\begin{itemize}
\item $0\leq r\leq 1/2$. The number $k$ of intervals
is equal to one (see eq.(\ref{kint1})). Equation (\ref{ME12}) yields directly
\begin{eqnarray*}
P(x,t) & \equiv & e^{-rt}\\
\theta (t) & = & 1-e^{-rt}
\end{eqnarray*}
which is, as expected, the exact result for the deposition of monomers
on a lattice, up to a factor $r$ in the flux corresponding to the target
area for each deposition.

\item $r=1$.  The two rate equations (\ref{ME12},\ref{ME22}) become identical
and reduce to the
known car-parking equation under the {\em ansatz}
$\displaystyle{P(x,t)=e^{-xt}\; q(t)}$.

\item $3/2\leq r\leq 2$. There is again only one interval.
{}From the rate equation we get directly
$$P(x,t)\equiv u_0(t) =\exp \left[e^{-(2-r)t}-(2-r)t-1\right]$$
and
$$\theta (t)=(2-r)\int_0^t dt'\; e^{(2-r)t'}\;  u_0^2(t') =
\frac{1}{2}\left[1-\exp [-2+2e^{-(2-r)t}]\right]$$
which is exactly the lattice dimer solution, up to a factor 1/2 due to
our definition of $\theta$.

\end{itemize}

\subsection{The case $1/2< r\leq 2/3$}
According to eq.(\ref{kint1}), this is the interval (for $r<1$) corresponding
to the lowest non trivial value $k=2$. There are only two reduced
probabilities, $q(t)$
and $P_2(x,t)$.  The function $\rho_2(t,t')$ of eq.(\ref{qt}) is given by
$$\rho_2(t,t')=\frac{e^{(2r-1)(t-t')}-1}{t-t'}-(2r-1) $$
and $K_2(x,t|t')$ of eq.(\ref{pl}) by
$$K_2(x,t|t')=e^{-rt'}\; \frac{e^{-x(t-t')}-e^{-(1-r)(t-t')}}{t-t'} $$
(see the appendix). After some elementary algebra based on
the integral representation eq.(\ref{pl}), one gets
from eq.(\ref{thet1}) a very simple expression for $\theta$~:
\begin{equation}
\theta =1-q^2(t)\;e^{rt} \label{theta2}
\end{equation}
We obtain the exact asymptotic behaviour of the coverage in the
following way.

{}From the inequality $0\leq q(t) \leq 1$ and equation (\ref{qt}), it follows
that $\displaystyle{q(t)<C\frac{e^{-(1-r)t}}{t}}$
which implies that the functions
$$
C_n(x,t) = \int_0^t e^{-(r-x)t'}\; q(t')\; t'^n\; dt'
$$
have a positive, finite, $t\rightarrow\infty$ limit for all $n\geq 0$ and
for all $x\in[0,r]$, that we denote $C_n(x)$. Expanding the denominator
of the kernel $\rho_2(t,t')$
in power of $t'/t$, we get from eq.(\ref{qt}) the asymptotic expansion
of $q(t)$ at large $t$
\begin{equation}
q(t) = \frac{e^{-(1-r)t}}{t}\left[ C_0(1-r)+\frac{C_1(1-r)}{t}
+O(\frac{1}{t^2})\right]        \label{qtas}
\end{equation}
Inserting this result in eq.(\ref{theta2}), we obtain for the coverage
 the asymptotic value $\theta (t=\infty )=1$, as expected, with
the approach~:
\begin{equation}
\theta (t)=1-C_0^2(1-r)\; \frac{e^{-(2-3r)t}}{t^2}+\cdots \label{th2}
\vspace{0.5cm}
\end{equation}
This equation shows that the kinetics remains exponentially driven over
the open interval $1/2 < r< 2/3$ but becomes $1/t^2$ at the end
point $r=2/3$.

\subsection{The case $2/3 < r < 3/4$}
We have explicitly studied this case, where three $x$-intervals $(k=3)$ are
involved, since we expect it to be typical of what happens over the
remaining range $r<1$.\\

Following the method of the previous section, we can derive for the
reduced probabilities $P_2$ and $P_3$ the following large-time
behaviour
\begin{eqnarray*}
P_2(x,t) & = & C_0(x,t)\;\frac{e^{-xt}}{t}\left[1+O(\frac{1}{t^2})\right]\\
P_3(x,t) & = & P_2(x,t)+O\left(\frac{e^{-(2r-1)t}}{t^2}\right)
\end{eqnarray*}
whereas the asymptotic behaviour of $q(t)$ is still given by eq.(\ref{qtas})
in spite of its being defined from a different kernel $\rho_3$.\\
The time derivative of $\theta$ may be expressed in terms of the reduced
probabilities as follows
$$
\dot{\theta}(t)=e^{rt}\left\{\int_{1-r}^{2r-1}P_2(x,t)\; P_2(r-x,t)\; dx
+2q(t)\;\int_{2r-1}^{2(1-r)} P_2(x,t)\; dx
+2q(t)\;\int_{2(1-r)}^r P_3(x,t)\; dx\right\}
$$
Inserting in this expression the previous asymptotic expansion, we get
\begin{equation}
\dot{\theta}(t) = \frac{A}{t^2}+\frac{2B}{t^3}  \label{th3}
\end{equation}
where $A$ and $B$ are constants depending only on $r$. In particular
$$A=\lim_{t\rightarrow\infty}\int_{1-r}^{2r-1} C_0(x,t)\; C_0(r-x,t)\; dx$$
We see that, except at  $r=2/3$ where $A$ vanishes, the leading behaviour
of $\theta$ is of the form
$$\theta (t) = \theta (\infty ) -\frac{A}{t}  $$
We expect this behaviour to hold over the whole range up to $r=1$ where
it is proved. We have checked this assumption numerically for $r=0.70$
and $r=0.9$.

For $r=2/3$ we recover the result of the previous section
$$\theta (t) = \theta (\infty )-\frac{B}{t^2} $$
which confirms that this unusual behaviour occurs only at this special
value of $r$.

\subsection{The case $4/3\leq r < 3/2$}
As in the previous section, this interval corresponds to the first
non-trivial one where $k=2$ in the case $r>1$. The kernels $K_2$ and
$\rho_2$ are defined by
\begin{eqnarray*}
K_2(x,t|t') & = & e^{-t'}\; e^{(r-1)t}\;
\frac{e^{-x(t-t')}-e^{-(2-r)(t-t')}}{t-t'}\\
\rho_2(t,t') & = & e^{-(2-r)t'}\;\int_0^{t-t'}
\frac{e^{-(3-2r)t''}[1+(3-2r)t'']-1}{t''^2}\;
\exp\left[ 1-e^{-(2-r)(t'+t'')}\right]\;dt''
\end{eqnarray*}
The expression for $\theta$ in terms of the reduced probability $q(t)$ is
obtained from eq.(\ref{thet2})~:
\begin{equation}
\theta =1-q^2(t)\;e^{2(2-r)t}
-(2-r)\int_0^t e^{(2-r)t'}\; q^2(t')\; dt' \label{thet}
\end{equation}
One observes that $\rho_2$ is negative, which implies the bound
$0< q(t) \leq u_0(t)$ and, from the mean-value theorem, the estimate
$$
\rho_2(t,t') = e^{-(2-r)t'}\;G(t-t')\;\left\{
\frac{1-e^{-(3-2r)(t-t')}}{t-t'}-(3-2r)\right\}
$$
where $\displaystyle{1\leq \exp (1-e^{-(2-r)t'})\leq G(t-t') \leq
\exp (1-e^{-(2-r)t})< e=2.718...}$.
Inserting this result in eq.(\ref{qt}) leads to the asymptotic time
expansion of $q(t)$
$$ q(t)=e^{-(2-r)t}\;\left\{ A(r)+\frac{B(r)}{t}+\cdots\right\} $$
where the bracket reduces to a constant  when $r$ goes to 3/2.
This shows from
eq.(\ref{thet}) that the jamming limit $\theta(\infty )$ is reached
as $1/t$.

\section{Summary and conclusions}

We have studied a one-dimensional model of random sequential adsorption
with discrete and continuous degrees of freedom, which depends
on one length parameter $r$. When this parameter varies over its range
$0< r < 2$ the model goes through the following regimes~:
\begin{itemize}
\item $0< r\leq 1/2$\\
monomer on a lattice; $\theta (t) = 1-e^{-rt}$
\item $1/2\leq r< 2/3$\\
total coverage; non trivial lattice dynamics~:
$\theta (t) = 1-Ae^{-(2-3r)t}/t^2$ for large $t$.
\item $r= 2/3$\\
total coverage; anomalous continuous dynamics~:
$\theta (t) = 1-A/t^2$ for large $t$.
\item $2/3< r< 3/4$\\
non-trivial asymptotic coverage; normal continuous dynamics~:
$\theta (t) =\theta (\infty )-A/t$ for large~$t$.
\item $r=1$\\
``car-parking" problem~: $\theta (t) =0.7476\cdots-A/t$ for large~$t$.
\item $4/3\leq r \leq 3/2$\\
same as for $2/3< r< 3/4$.
\item $3/2\leq r< 2$\\
lattice dimer model;
$\displaystyle{\theta (t)=\frac{1}{2}\left[1-\exp [-2+2e^{-(2-r)t}]\right]}$
\item $3/4< r< 4/3$\\
Although we have not investigated this interval analytically
(except for $r=1$),
we expect the same regime as in the bordering intervals $2/3< r< 3/4$
and $4/3\leq r \leq 3/2$ to hold, that is a non trivial asymptotic coverage
decreasing with $r$ and a $1/t$ normal continuous dynamics. We have checked
this conjecture numerically.
\end{itemize}
The remarkable point is that the kinetics of the model exhibits three
``phases"~: for $0\leq r<2/3$ where it is lattice-like, for $2/3<r <3/2$
where it is continuous-like and for $3/2\leq r\leq 2$ where it is
lattice-like again. At the transition point $r=2/3$ it becomes ``anomalous"
since the jamming limit is approached in $1/t^2$ in contrast with the
general belief  that the exponent $n$ of the power-law decay is equal to
the inverse of the number of degrees of freedom per particle.

The regime on both sides of the transition is characterised by a same
typical cross-over time\cite{VTRT}
 $\tau = \frac{1}{|2-3r|}$, defined from the slope
of the exponential in eq.(\ref{th2}) or from the ratio $B/A$ in
eq.(\ref{th3}). This time is such that for $t\ll \tau$ the dynamics is
dominated by a $1/t^2$ behaviour in the two ``phases"; only for $t\gg\tau$
does the characteristic long time behaviour, exponential for $r<2/3$,
in $1/t$ for $r>2/3$, emerge. Since $\tau\rightarrow\infty$ when
$r\rightarrow 2/3$, this long time regime is squeezed at the transition
point $r=2/3$, leaving only the $1/t^2$ behaviour.

To find for $\theta (t)$ an expression in a closed form is a very difficult
technical problem, comparable to the determination of the correlation
function in the standard ``car-parking" model\cite{Bonnier}. However,
the properties of the kernel $\rho_k$ allow us to obtain an iterative solution
of eq.(\ref{qt}) from which approximate expressions for the long time
coverage can be obtained.

Finally, it may be interesting to investigate these kinds of models in a
more realistic physical context, such as a two dimensional substrate,
a disordered distribution of adsorbing sites or the possibility for
a particle to overlap several sites. It is
presumably difficult to get an analytical insight in such models, but their
properties can be numerically analysed using as a guide the
one dimensional results.\vspace{1cm} \\

\noindent{{\bf Aknowlegements}}\\
We thank G. Tarjus for discussions and for pointing out to us that
model (I) and model (II) are equivalent.

\newpage
\appendix
\centerline{{\Large{\bf APPENDIX}}}

\section{The master equations}
In this appendix, we derive the master equations
for the probability of finding at time $t$ an unoccupied interval ({\em gap})
of given length. Let $P_n(x,y,t)$ be the probability for finding a gap
of length at least $x+y+n-1$, where $n$ is the number of adsorbing sites
in the gap and $x(y)$ the distance between the left (right) edge of the gap
and the  last left(right) site in the gap
(see figure 4). $P_1(0,0,t)$ is the probability for finding a gap containing
at least one site and its complement,
 $1-P_1(0,0,t)$, defines the probability that a
site is covered by a particle, which is equivalent to the occupancy rate~:
$$ \theta (t) = 1-P_1(0,0,t)$$
The rate equation for this quantity is (in these equations and in the
subsequent rate equations we set the effective flux of particles to unity)~
\begin{eqnarray}
-\frac{\partial P_1(0,0,t)}{\partial t} & = &
\int_0^r P_1(x,r-x,t)\; dx\qquad {\rm for}\; r\in [0,1]\label{P11}\\
& = & \int_{r-1}^1 P_1(x,r-x,t)\; dx \qquad {\rm for}\; r\in [1,2]
\label{P12}
\end{eqnarray}
Therefore to determine $\theta$ one needs to know $P_1(x,y,t)$ only
for $x$ and $y$ less than $r$ (case $r\leq 1$) or greater than $r-1$
(case $r\geq 1$)
and set $x+y=r$. The general rate equations are obtained as
usual, by counting the different ways of destroying a gap~:
\begin{eqnarray}
{\rm case}\; r\leq 1 && x\in[0,r]\quad y\in[0,r]\quad n\geq 2\nonumber\\
& & \nonumber\\
&&-\frac{\partial P_n(x,y,t)}{\partial t}=\label{ME1}\\
&&[x+y+(n-2)r]P_n(x,y,t)+\int_x^r P_n(x',y,t)dx'+\int_y^r P_n(x,y',t)dy'
\nonumber\\
 & & +u(x+r-1)\int_0^{x+r-1}P_{n+1}(x',y,t)dx'+
     u(y+r-1)\int_0^{y+r-1}P_{n+1}(x,y',t)dy' \nonumber\\
 & & \nonumber\\
{\rm case}\; r\geq 1 && x\in[r-1,1]\quad y\in[r-1,1]\quad n\geq 2\nonumber\\
& & \nonumber\\
&&-\frac{\partial P_n(x,y,t)}{\partial t}=
 [(x-r+1)+(y-r+1)+(n-2)(2-r)]P_n(x,y,t)\nonumber\\
 & & +\int_x^1 P_n(x',y,t)dx'+\int_y^1 P_n(x,y',t)dy'\nonumber\\
 & & +\int_{r-1}^{b(x)}P_{n+1}(x',y,t)dx'+
     \int_{r-1}^{b(y)}P_{n+1}(x,y',t)dy'  \label{ME2}
\end{eqnarray}
where $u(t)$ is the Heaviside step function and,
in the last equation, $b(x)={\rm Inf}\; \{x+r-1,1\}$.
The probability $P_n(x,y,t)$ is subject to the initial condition
$$ P_n(x,y,t=0)=1$$
{}From these
equations, we see that for $n\geq 2$, the $n$
dependence can be factorised out~:
\begin{eqnarray}
P_n(x,y,t) & = & e^{-(n-2)rt}\;P_2(x,y,t)\qquad {\rm for}\; r\in [0,1]
\label{factorN1}\\
& = & e^{-(n-2)(2-r)t}\;P_2(x,y,t)\qquad {\rm for}\; r\in [1,2]
\label{factorN2}
\end{eqnarray}
The function $P_2(x,y,t)$ must be symmetric in its spatial arguments.
Moreover,  for $n\geq 2$ the left and the right parts of the gap cannot
be simultaneously affected by the deposition of one dimer, hence the
$x$ and $y$ dependence are uncorrelated. We deduce from these considerations
that~:
\begin{equation}
P_2(x,y,t)=P(x,t)\; P(y,t) \label{factorN3}
\end{equation}
reducing the problem to finding one unknown function, $P(x,t)$, which we call
in
the following, a {\em reduced probability}. \\
The rate equation for the function $P_1(x,y,t)$ is different from
eqs(\ref{ME1},\ref{ME2}) and depends on the value of $x+y$ with respect to $r$.
However, the arguments leading to the factorisation
of the $x$ and $y$ dependences and of the $n$ dependence remain valid
for $x+y\geq r$ which is precisely the region of interest, leading to~:
$$P_1(x,y,t)=e^{rt}\;P(x,t)\; P(y,t) \qquad x+y\geq r$$
{}From this factorisation property and
the rate equations(\ref{P11},\ref{P12}),
we can  express the occupancy rate in terms of the function $P$,
leading to eqs.(\ref{thet1},\ref{thet2})~:
\begin{eqnarray*}
\theta (t) & = & \int_0^t dt'\; e^{rt'}\int_0^r dx\; P(x,t')\;P(r-x,t')
\qquad {\rm for}\; r\in [0,1] \\
      & = & \int_0^t dt'\; e^{(2-r)t'}\int_{r-1}^1 dx\; P(x,t')\;P(r-x,t')
\qquad {\rm for}\; r\in [1,2]
\end{eqnarray*}

{}From eqs.(\ref{ME1},\ref{ME2}) and the factorisation properties of
$P_n(x,y,t)$, eqs.(\ref{factorN1},\ref{factorN2},\ref{factorN3})
we deduce the rate equations for the
function $P$, eqs.(\ref{ME12},\ref{ME22}) of section 3.1~:
\begin{eqnarray}
{\rm case}\; r\leq 1&&x\in[0,r]\nonumber\\
& & \nonumber\\
-\frac{\partial P(x,t)}{\partial t} & = &
xP(x,t)+\int_x^r P(x',t)dx'+
u(x+r-1)\; e^{-rt}\int_0^{x+r-1}P(x',t)dx' \label{rate12}\\
 & & \nonumber\\
{\rm case}\; r\geq 1&&x\in[r-1,1]\nonumber\\
& & \nonumber\\
-\frac{\partial P(x,t)}{\partial t} & = &
(x-r+1)P(x,t)+\int_x^1 P(x',t)dx'+ e^{-(2-r)t}\;
\int_{r-1}^{b(x)}P(x',t)dx' \label{rate22}
\end{eqnarray}
Due to the last term in eqs.(\ref{rate12},\ref{rate22}), $P(x,t)$ appears
to be piecewise
defined with respect to $x$, in successive intervals of length $(1-r)$
within the range $(0,r)$ for $r\leq 1$, or of length $(r-1)$
within the range $(r-1,2-r)$ for $r\geq 1$. To be more explicit, let us
define the set of intervals $I_\ell$ in the following way~:\\

\noindent For $0\leq r<1$ define $k\geq 1$ such that
$\displaystyle{\frac{k-1}{k}<r\leq\frac{k}{k+1}}$; then~:
\begin{eqnarray*}
&&I_\ell =\left[(\ell -1)(1-r),\ell (1-r)\right] \qquad 1\leq \ell \leq k-1\\
&&I_k=\left[(k-1)(1-r),r\right] \\
{\rm which~are~such~that}&& \bigcup_{\ell =1}^k I_{\ell}=[0,r]
\end{eqnarray*}

\noindent For $1<r\leq 2$ define $k\geq 1$ such that
$\displaystyle{\frac{k+2}{k+1}<r\leq\frac{k+1}{k}}$; then~:
\begin{eqnarray*}
&&I_1=\left[(2-r),1\right]\\
{\rm only~for }\; k\geq 2&&I_2=\left[(k-1)(r-1),(2-r)\right]\\
{\rm only~for }\; k\geq 3&&
I_\ell =\left[(k-\ell +1) (r-1),(k-\ell +2) (r-1)\right]
\qquad 3\leq \ell \leq k \\
{\rm which~are~such~that}&& \bigcup_{\ell =1}^k I_{\ell}=[r-1,1]
\end{eqnarray*}

The number $k$ of such intervals is directly expressed in terms of $r$~:
$$
k = \left[\frac{r}{1-r}\right]+1\quad  {\rm for}\; 0\leq r< 1
\qquad
k = \left[\frac{2-r}{r-1}\right]+1\quad  {\rm for}\; 1< r\leq 2
$$
where $[X]$ means integer part of $X$. In the following,
for a given value of $r$, we will denote by
$P_{\ell}(x,t),\quad \ell=1,...,k$ the restriction of $P(x,t)$ to the
interval $I_\ell$.

\section{Integral equations for the generic case $2/3<r<3/2$,
 $r\neq 1$}
In this appendix, we derive the representation
 of eqs.(\ref{pl},\ref{qt})
from eqs(\ref{rate12},\ref{rate22}) and we give the expression of the kernels
$K_\ell$ and $\rho_k$ and of the function $u_0$.\\

We first remark that the reduced probability $P_1(x,t)$ is in fact
independent of $x$~: $P_1(x,t)\equiv q(t)$.
Consider the left bordering site of the gap, in the case $r\leq 1$.
There is a vacant space of length {\em at least} $1-r$ from its right
edge independent on the occupation of its left neighbour;
 this means that $P(x,t)$ is independent of $x$ for $x\in[0,1-r]$ .
The same property holds, in the case $r\geq 1$, for $x\in [2-r,1]$.\\

We start with the case $r<1$. For $1-r\leq x\leq r$, we derive
eq.(\ref{rate12}) with respect to $x$ to get
$$ -\frac{\partial^2 P(x,t)}{\partial x\partial t}=
x\frac{\partial P(x,t)}{\partial x}+
 e^{-rt}P(x+r-1,t)$$
Then integrating with respect to $t$, using the initial condition
$\displaystyle{ \frac{\partial P(x,0)}{\partial x}=0}$,
we obtain
$$ \frac{\partial P(x,t)}{\partial x}=-\int_0^t
 e^{-rt'-x(t-t')}\;P(x+r-1,t')\; dt'$$
Finally, we integrate with respect to $x$, taking into account the
boundary condition $P(x=1-r,t)=q(t)$, to get
\begin{equation}
P(x,t)=q(t)-e^{-(1-r)t}\;\int_0^t dt'\; e^{-(2r-1)t'}\; \int_0^{x+r-1} dx'\;
e^{x'(t'-t)}\; P(x',t')         \label{plo}
\end{equation}
By using repeatedly eq.(\ref{plo})
as $x$ increases, one obtains equation (\ref{pl})~:
$$
P_{\ell}(x,t) = q(t)+\int_0^t K_{\ell}(x,t|t')\; q(t')\; dt \qquad
2\leq \ell \leq k
$$

For $x\in I_2$ then $x'\in I_1$ where
$P(x',t')=q(t')$ and eq.(\ref{plo}) gives eq.(\ref{pl}) with $\ell =2$,
and the following expression for the kernel $K_2$
\begin{equation}
K_2(x,t|t')=e^{-rt'}\; \frac{e^{-x(t-t')}-e^{-(1-r)(t-t')}}{t-t'} \qquad
r<1 \label{K2}
\end{equation}

When $x\in I_3$, by including the previous result in eq.(\ref{plo}), we
get eq.(\ref{pl}) for $\ell =3$ with the explicit form of the kernel $K_3$,
and so on.\\

Analogous manipulations can be performed on eq.(\ref{rate22}) in the case
$r>1$ for $x\leq 2-r$, where the boundary condition is now
$P(x=2-r,t)=q(t)$.The equation equivalent to eq.(\ref{plo}) is
\begin{equation}
P(x,t)=q(t)+e^{2(r-1)t}\;\int_0^t dt'\; e^{-rt'}\; \int_{x+r-1}^1 dx'\;
e^{x'(t'-t)}\; P(x',t')         \label{pl1}
\end{equation}
which, by repeated application as $x$ decreases, gives the representation
of eq.(\ref{pl}). For $x\in I_2$, one gets the kernel $K_2$
\begin{equation}
K_2(x,t|t')=e^{-t'+(r-1)t}\; \frac{e^{-x(t-t')}-e^{-(2-r)(t-t')}}{t-t'}
\qquad r>1 \vspace{1cm} \label{K22}
\end{equation}

Assuming that, for a given value of $r$ all the $K_\ell$ are known, one
can define a kernel $\sigma_k (t,t')$
$$\sigma_k (t,t')=\sum_{\ell =2}^k\int_{I_\ell}dx\; K_\ell (x,t|t')$$
and derive an integral equation for $q(t)$. Considering eqs.(\ref{rate12},
\ref{rate22}) where $x$ is fixed to the value $x=1-r$ (case $r<1$) or
$x=2-r$ (case $r>1$) and using eq.(\ref{pl}) for $P_\ell (x,t)$, one gets
respectively~:\\
\begin{eqnarray*}
r<1 \qquad -\frac{d q}{d t}(t) & = &
rq(t)+\int_0^tq(t')\;\sigma_k (t,t')\; dt'\\
r>1 \qquad -\frac{d q}{d t}(t) & = &
(2-r)[1+e^{-(2-r)t}]q(t) +e^{-(2-r)t}\int_0^tq(t')\; \sigma_k (t,t')\; dt'
\end{eqnarray*}
By integrating with respect to $t$ with the initial condition $q(0)=1$,
we get the equation (\ref{qt})
$$
q(t) = u_0(t)\left\{ 1+\int_0^t \rho_k (t,t')\; q(t')\; dt'\right\}
$$
with
$$
\begin{array}{lcl}
r<1 \qquad u_0(t)=e^{-rt} & , &
\displaystyle{\rho_k(t,t')=\int_t^{t'}e^{rt''}\;\sigma_k(t'',t')\; dt''}\\
r>1 \qquad u_0(t)= \exp\left[e^{-(2-r)t}-(2-r)t-1\right] & , &
\displaystyle{\rho_k(t,t')=\int_t^{t'}
\exp\left[1-e^{-(2-r)t''}\right]\;\sigma_k(t'',t')\; dt''}
\end{array}
$$

\newpage
\newcommand{\refer}[4]{{\em #1} {\bf #2},#3 (#4)}

\newpage
{\Large {\bf Figure captions}}\\
\begin{itemize}
\item[]{\bf Figure 1} (a) description and parametrization of model (I)~:
deposition of dimers on extended adsorbing sites. (b)
description and parametrization of model (II)~: deposition of segments
on localised adsorbing sites.

\item[]{\bf Figure 2} The jamming limit of the model as a function of
the length $r$ of the adsorbed segments. The values of $r$ separating
the different regimes are indicated.

\item[]{\bf Figure 3}  Two extreme situations for the deposition in
model (II).

\item[]{\bf Figure 4} The parametrisation of the gap.

\end{itemize}


\begin{thebibliography}{99}
    \bibitem{Evans} For a comprehensive review of the theoretical aspects
and experimental applications of RSA, see
J.W. Evans,{\em Rev. Mod. Phys.} {\bf 65}, 1281 (1993)
    \bibitem{EvansNord}
J.W. Evans and R.S. Nord,{\em J. Stat. Phys.}{\bf 38}, 681 (1985)
    \bibitem{Feder}J. Feder {\em J. Theor. Biol.} {\bf 87}, 237 (1980)
    \bibitem{Pomeau}Y. Pomeau, \refer{J. Phys. A}{13}{L193}{1980}
    \bibitem{Swendsen} R. Swendsen, \refer{Phys. Rev. A}{24}{504}{1981}
    \bibitem{Sherwood}J. D. Sherwood, {\em J. Phys. A} {\bf 23}, 2827 (1990)
    \bibitem{Viot}P. Viot and G. Tarjus
{\em Europhys. Lett.} {\bf 13}, 295 (1990)
    \bibitem{Vigil}R.D. Vigil and R. B. Ziff {\em J. Chem.
Phys.} {\bf 91}, 2599 (1989)
    \bibitem{Vigil2}R.M. Ziff and R.D.Vigil,
{\em J. Phys.}{\bf A23}, 5103 (1990)
    \bibitem{Ricci1}P. Viot, G. Tarjus, S.M. Ricci and J. Talbot
{\em J. Chem. Phys.} {\bf 97}, 5219 (1992)
   \bibitem{Ricci}P. Viot, G. Tarjus, S.M. Ricci and J. Talbot
{\em Physica A} {\bf 191}, 248 (1992)
    \bibitem{Renyi} A. Renyi, {\em Publ. Math.
Inst. Hung. Acad. Sci.}{3}{109}{1958}
    \bibitem{japs} S. Hosaka, H Koyanagi, A. Kikukawa,Y. Maruyama and
R. Imura, Proceedings of the Int. Conf. on STM, 1993.
    \bibitem{Aime} J.P. Aim\'e (private communication).
    \bibitem{Tarjus2} X. Jin, N.-H. L. Wang, G. Tarjus and J. Talbot
\refer{J. Phys. Chem}{97}{4256}{1993}
    \bibitem{Flory}P.J. Flory, {\em J. Am. Chem. Soc.}{61}{1518}{1939}
    \bibitem{VTRT} A similar cross-over effect has been observed
in the RSA kinetics of very elongated particles in ref.\cite{Ricci}
    \bibitem{Bonnier} B. Bonnier, D. Boyer and P. Viot, {\em J. Phys. A}
{\bf 27}, 3671 (1994)

\end{thebibliography}
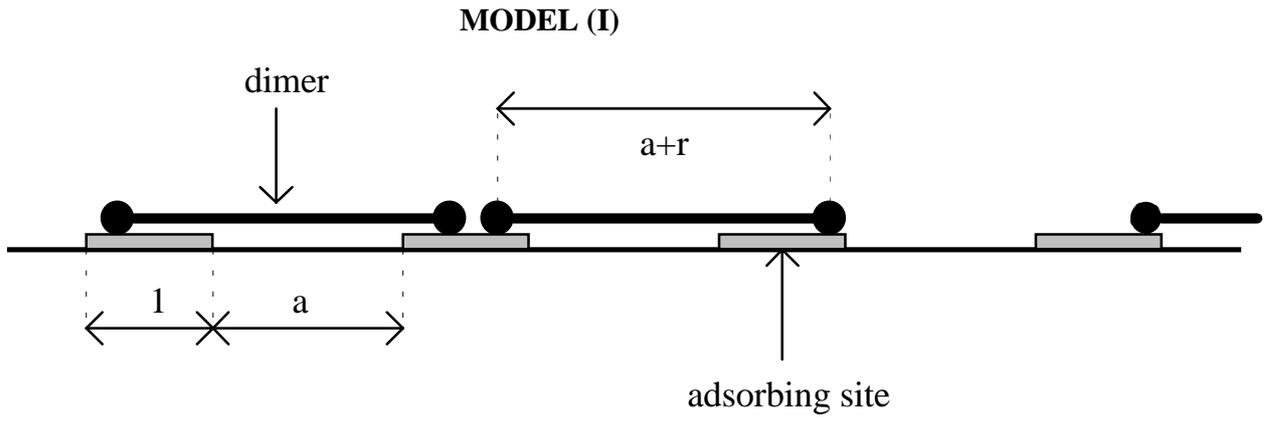
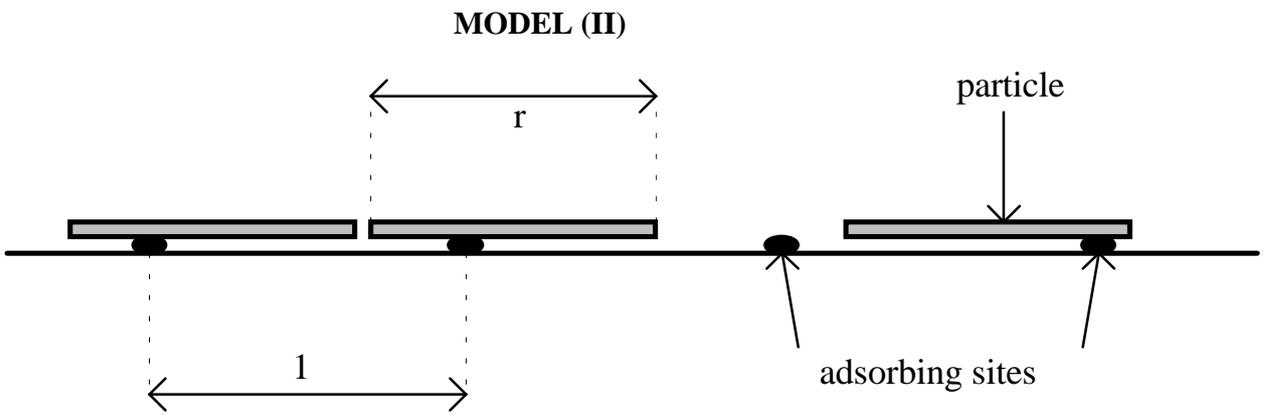
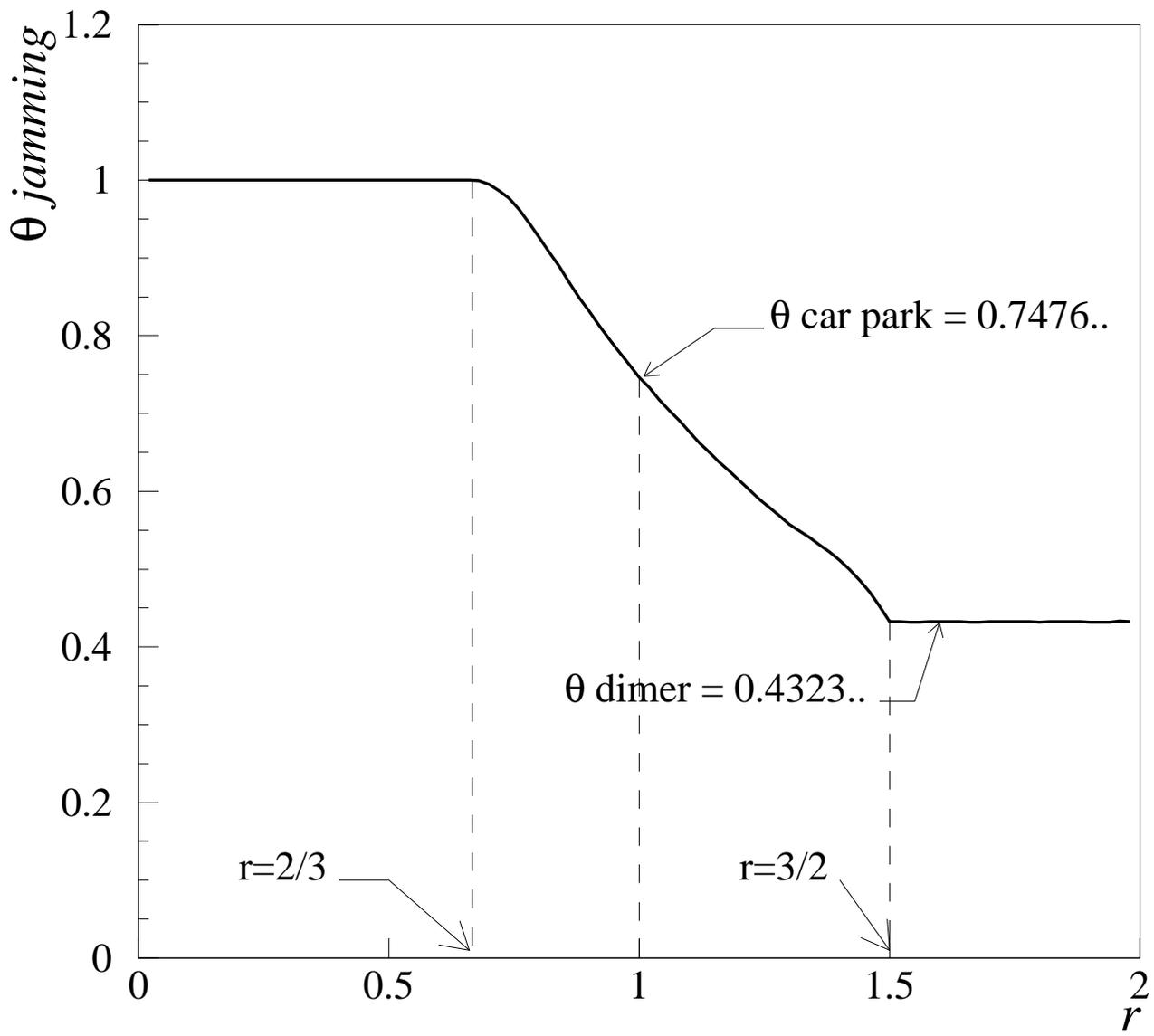
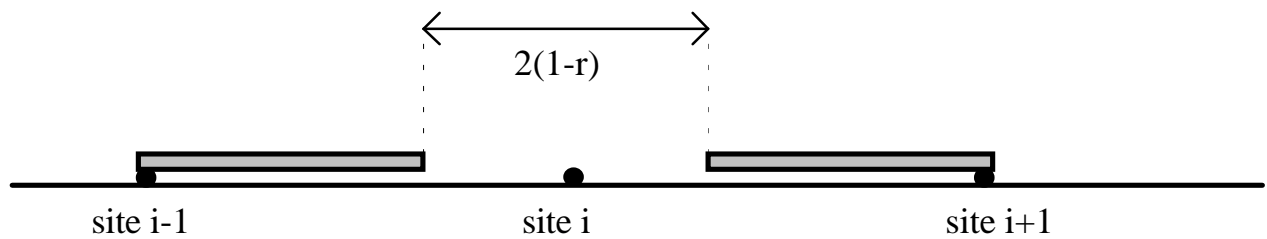
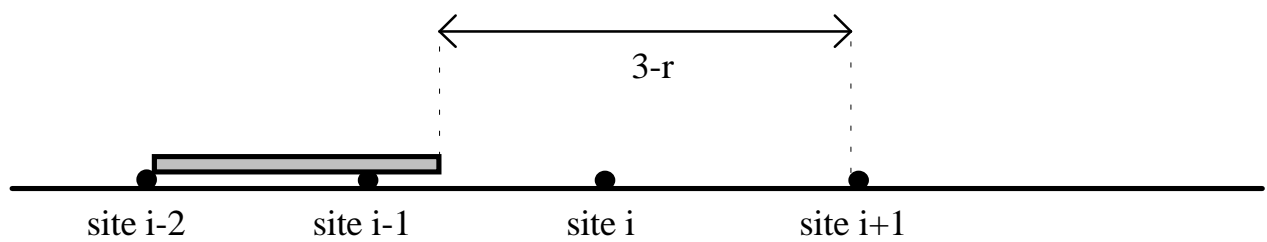
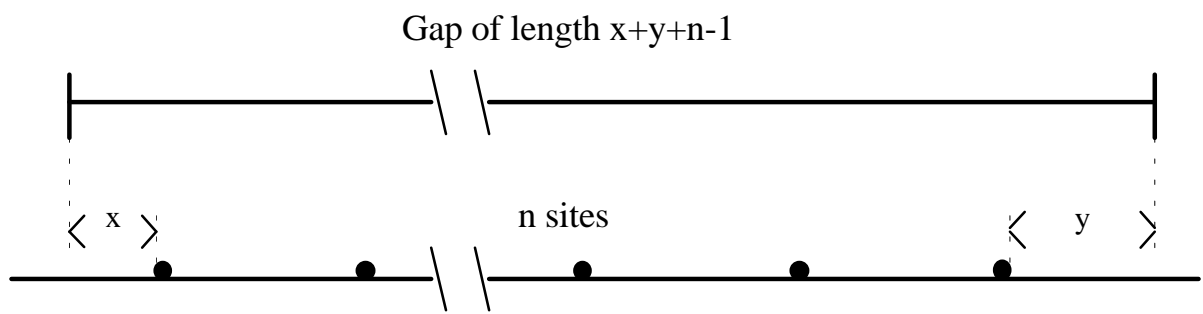
\end{document}